\documentclass[journal=jacsat,manuscript=article]{achemso}

\usepackage{chemformula} 
\usepackage[T1]{fontenc} 

\author{Bruno H. S. Mendon\c{c}a}
\affiliation[UFMG]{Departamento de F{\'i}sica, ICEX, Universidade Federal de Minas Gerais, CP 702, Belo Horizonte 30123-970, MG, Brazil}
\email{brunnohennrique13@gmail.com}

\author{Elizane E. de Moraes}
\affiliation[UFBA]{Instituto de F{\'i}sica, Universidade Federal da Bahia, Campus Universit{\'a}rio de Ondina, Salvador 40210-340, BA, Brazil}

\author{H{\'e}lio Chacham}
\affiliation[UFMG]{Departamento de F{\'i}sica, ICEX, Universidade Federal de Minas Gerais, CP 702, Belo Horizonte 30123-970, MG, Brazil}

\title[An \textsf{achemso} demo]
  {Enhanced diffusion in self-nanoconfined water channels between periodically modulated surfaces: insights from molecular dynamics simulations}

\abbreviations{IR,NMR,UV}

\begin{document}







\begin{abstract}
Water nanoconfinement is known to occur inside material void spaces, such as 2D confinement between surfaces, 1D confinement inside nanotubes, and variable-dimension confinement inside nanoporous materials. In the present work we investigate, through molecular dynamics simulations, the morphologies and self-diffusion coefficient of water channels that are nanoconfined in the void space between adjacent surfaces of nanotube bundles – an existing class of materials. In our simulations, we begin with water filling completely the void space, and then we progressively increase the inter-surface separation, maintaining the water content. We find that, as the inter-surface separation progresses, the dimensionality of the water channel decreases from 2D to 1D, the latter consisting of self-confined water channels along surface grooves. The morphologies and self-diffusion coefficients of these 1D water nanochannels are strongly dependent on the nature of the water-surface interaction and on the diameter of the nanotubes. Interestingly, as we decrease the nanotube diameter from 10 to 5 nm, the self-diffusion coefficients of the 1D channels increase by tenfold for hydrophilic surfaces and by sixfold for hydrophobic surfaces, surpassing, in both cases, the bulk water values. We also investigated the water channels at the interstitial voids of the bulk bundle material, finding 1D water channels that are similar to the surface ones.
\end{abstract}

\section{Introduction}

Water nanoconfinement is known to occur inside several types of material void spaces. For example, ordered 2D confinement occurs in voids between surfaces \cite{geim2021exploring,qin2017high,chacham2020controlling,luo2008ferroelectric}, either ordered \cite{hummer2001water} or disordered \cite{luo2008ferroelectric} 1D confinement may occur inside nanotubes, and disordered, variable-dimension confinement may occur inside nanoporous materials \cite{mochizuki2015solid,yang2017hierarchically,ding2017two}. Such confinement types may lead to novel phenomena such as enhanced self-diffusion \cite{mendoncca2023water,mendoncca2020water,mendoncca2020water2,de2020water,mendoncca2019diffusion} and dielectric responses \cite{wang2024plane}, and percolation transitions \cite{pereira2022aerosol,mendoncca2024conduction,jha2016liquid}. These phenomena depend not only on water channel dimensionality, but also on the nature of the interaction between water and surface \cite{lenz2025influence,nair2012unimpeded}.

In the present work we investigate, through molecular dynamics simulations, the morphology and self-diffusion coefficient of nanoconfined water channels between surfaces of nanotube bundles. Such bundles have been observed for different nanotube materials \cite{fu2018review,colomer2004study,kingston2004efficient}, with nanotube diameters ranging from a few nanometers to more than 100 nm. The surfaces of nanotube bundles are ordered and periodically modulated along one direction, forming a sequence of half-nanotube surfaces with surface “grooves” between nanotubes. When exposed to gases, it has been predicted \cite{colomer2004study} and observed \cite{kingston2004efficient} that the gas molecules preferentially adsorb at the groove regions, forming 1D structures.

In our simulations, we begin with water filling completely the void space, and then we progressively increase the inter-surface separation, maintaining the water content. We find that, as the inter-surface separation progresses, the dimensionality of the water channel decreases from 2D to 1D, the latter consisting of self-confined water channels along the surface grooves. The morphologies and the self-diffusion coefficients of these 1D water nanochannels are strongly dependent on the nature of the water-surface interaction and on the diameter of the nanotubes. Particularly, as we decrease the nanotube diameter from 10 to 5 nm, the self-diffusion coefficients of the 1D channels increase by tenfold for hydrophilic surfaces, and by sixfold for hydrophobic surfaces, surpassing, in both cases, the bulk water values. We also investigate water channels at the interstitial void regions of the bulk bundle material, where each interstitial region has three adjacent ``groove'' regions, with threefold symmetry. We find that the interstitial water channels spontaneously break that symmetry and concentrate at one of the three groove regions, becoming similar to the surface ones.

This article is organized as follows: Section II presents the simulation methodology and the definition of the models. Section III discusses the main results, and Section IV summarizes the conclusions.

\section{Simulation methodology}

\begin{figure}[H]
	\begin{center}
		\includegraphics[width=5.4in]{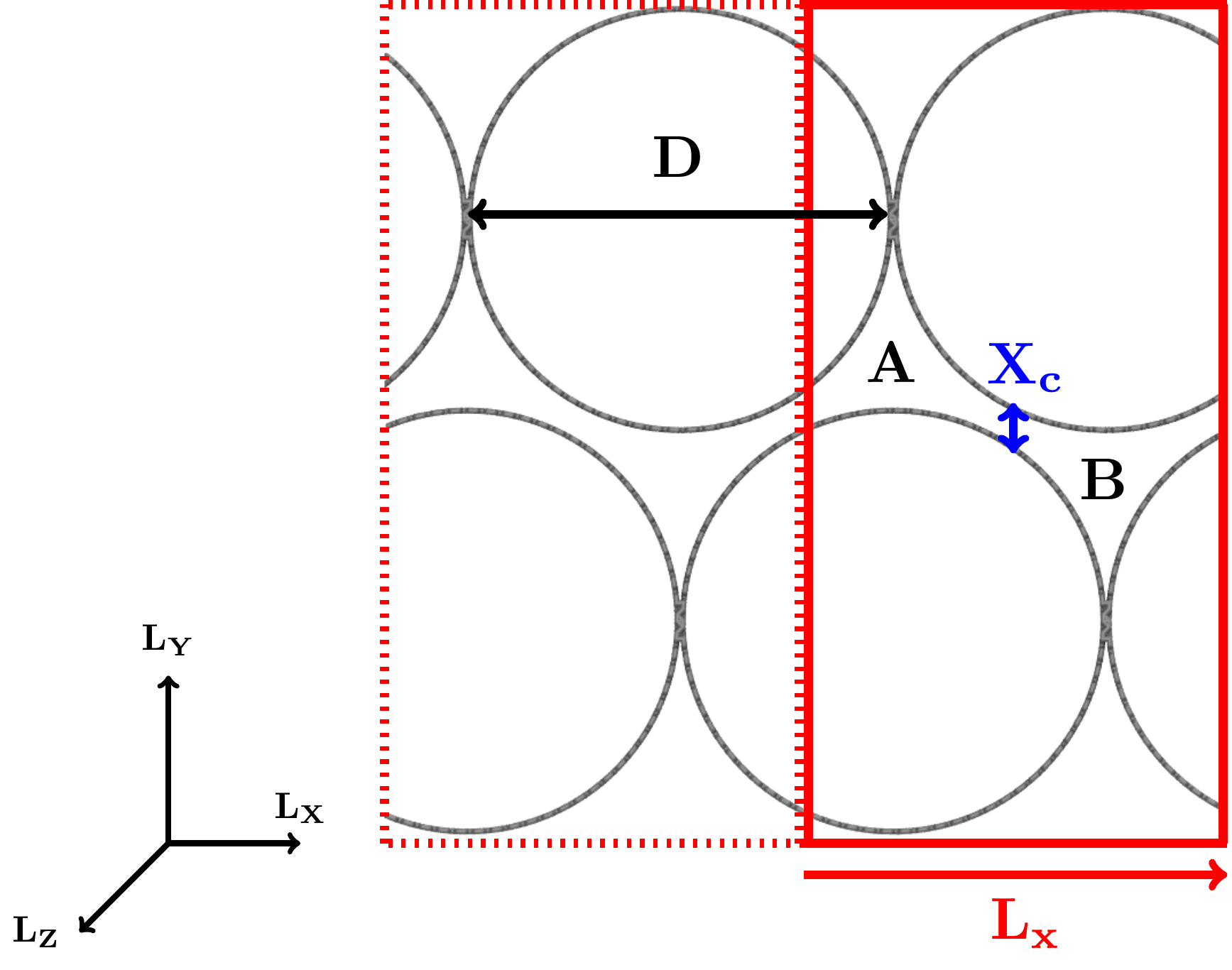}
	\end{center}
	\caption{Model of nanoconfined water between adjacent surfaces of nanotube bundles. The water molecules are confined to the nanochannel formed by regions A and B, with periodicity defined by the lattice vectors L$_{X}$ (horizontal in the figure) and L$_{Z}$ (perpendicular to the figure). The considered nanotube diameters D are 5, 10, 20, 40 nm, and the distance between the surfaces ( X$_{c}$, the channel width) varies from 1.0 to 3.0 nm.}     
	\label{fig_system}
\end{figure}

Molecular dynamics simulations with constant particle number N, volume V, and temperature T, were performed to analyze the dynamic behavior of water that is nanoconfined between adjacent  surfaces of
nanotube bundles, as shown in Figure~\ref{fig_system}.
The water model used was TIP4P/2005 \cite{abascal2005general}. The surfaces of the carbon nanotube bundles are represented by two
infinite, parallel rows of carbon nanotubes (CNTs) with diameters D that are placed at a distance d from each other, so that the minimum vertical distance X$_{c}$ between the surfaces is 0.5 nm, as indicated in Figure~\ref{fig_system}. This establishes 2D nanochannels in the void space between the surfaces, where the water molecules are inserted. The periodic boundary conditions are established with the horizontal lattice vector L$_{X}$ indicated and a lattice vector L$_{Z}$ perpendicular to the figure. The confinement originates nanochannels with adjacent regions labelled as A and B in the figure.

The carbon-carbon and carbon-water interactions were modeled using the Lennard-Jones (LJ) potential as follows. The classical potential for the interaction between carbon atoms was described with an energy of $\epsilon_{CC}=0.105067$~kcal/mol and an effective diameter of $\sigma_{CC}=3.851$~\AA~\cite{hummer2001water}. For carbon-oxygen interactions we used two distinct potentials, one hydrophilic and one hydrophobic with $\epsilon_{CO}=0.1142$~kcal/mol and $\sigma_{CO}=3.28$~\AA~(labeled as hydrophilic due to its attractive character to water) and reduced carbon-water interaction strength with $\epsilon_{CO}=0.0645$~kcal/mol and $\sigma_{CO}=3.41$~\AA~(i.e. hydrophobic), as done in previous work \cite{hummer2001water,moskowitz2014water}. Simulations were performed using the Large Scale Atomic/Molecular Massively Parallel Simulator (LAMMPS) package \cite{plimpton1995fast}. Periodic boundary conditions in the axial direction and a cutoff radius of 12~\AA~ in the interatomic potential were used. The water structure is constrained using the SHAKE algorithm \cite{ryckaert1977numerical}, with a tolerance of 10$^{-4}$. Long-range Coulomb interactions were computed using the Particle-Particle Particle-Mesh (PPPM) method \cite{ostler2017electropumping}. To prevent real charges from interacting with their own images, we set the dimensions of the simulation box perpendicular to the axial axis to 100~nm. The system temperature was maintained at 300~K, controlled by the Nosé–Hoover thermostat with a damping time of 100~fs \cite{nose2002molecular}. In all simulations, the nanosheets were kept rigid with zero velocity at the center of mass. This procedure has been used in several similar simulations and has proven to be a very reasonable approximation when compared to the case in which the thermostat is applied to the entire system \cite{kotsalis2004multiphase,hanasaki2006flow}. The systems were equilibrated for 5 ns and the properties of interest were calculated after this stage for another 5 ns, summing a total simulation time of 10 ns. The time step was 1 fs in all runs, and the properties were stored every 300 fs.

\section{Results and discussion}

\begin{figure}[H]
	\begin{center}
		\includegraphics[width=4.4in]{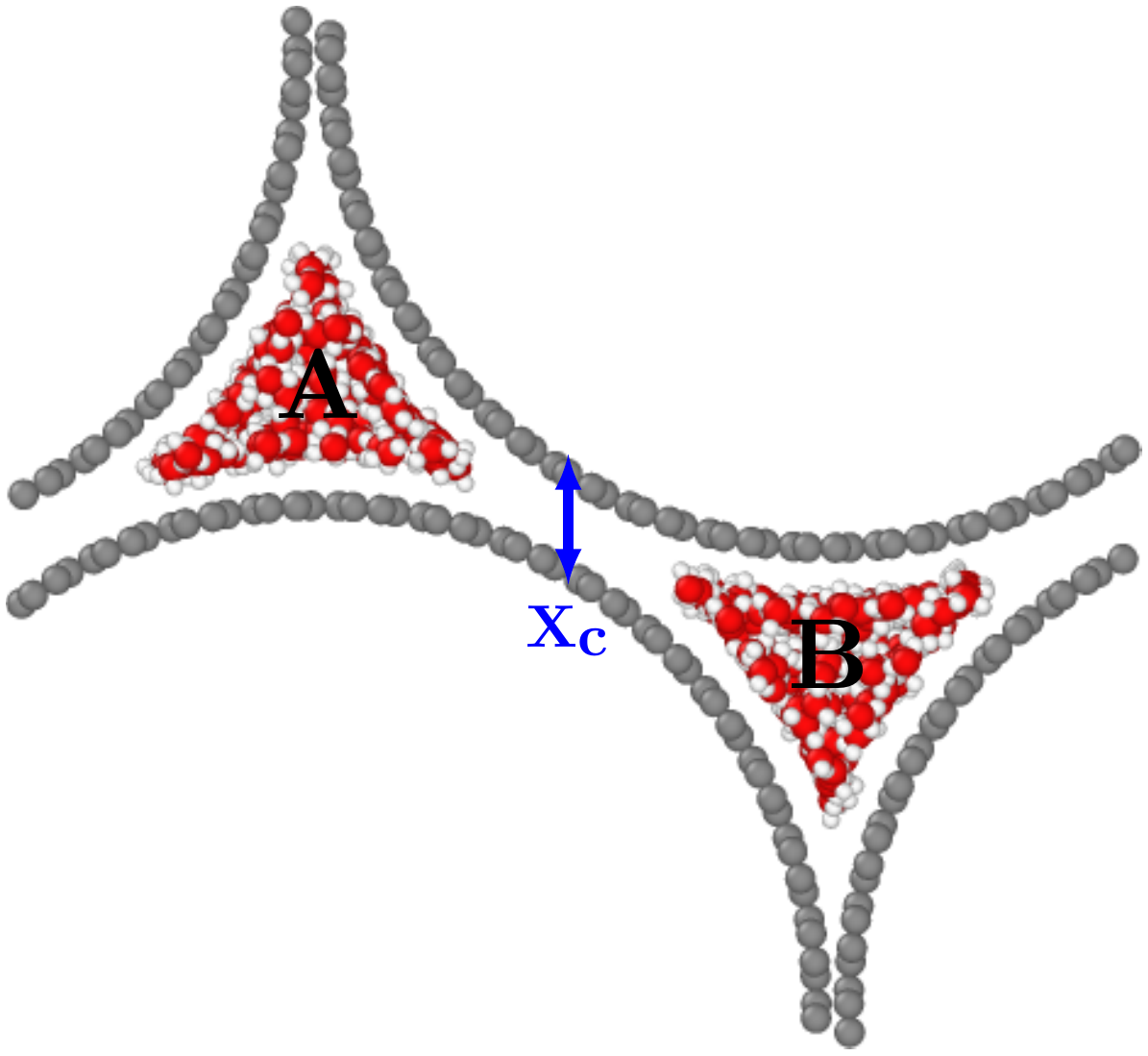}
	\end{center}
	\caption{Atomic scale snapshot of regions A and B of Figure \ref{fig_xc}. The minimum vertical distance between the rows of carbon nanotubes is X$_{c}$, as indicated. This creates a two-dimensional void region between the rows of carbon nanotubes, where water molecules are inserted at a density of 0.310 molecules/nm$^{3}$.}     
	\label{fig_xc}
\end{figure}

As we described in the previous section, we confine the water to channels within the void space between the parallel surfaces of two nanotube bundles. The channel thickness, that is, the distance between the adjacent surfaces, will be designated by the variable X$_{c}$ shown in Figure \ref{fig_xc}. We will consider X$_{c}$ values between 1.0 nm and 3.0 nm. Smaller diameters do not allow water passage, and larger diameters do not change water behavior after 3 nm. The density of water confined in each channel was 0.310 molecules/nm$^{3}$, and the nanotube diameters ranged from 5 nm to 10 nm. For larger diameters, the behavior of the water molecules is the same, considering the same density of confined water molecules. We also analyzed two types of interactions between the nanosheets and water molecules: hydrophobic and hydrophilic.

\begin{figure}[H]
	\begin{center}
		\includegraphics[width=6.4in]{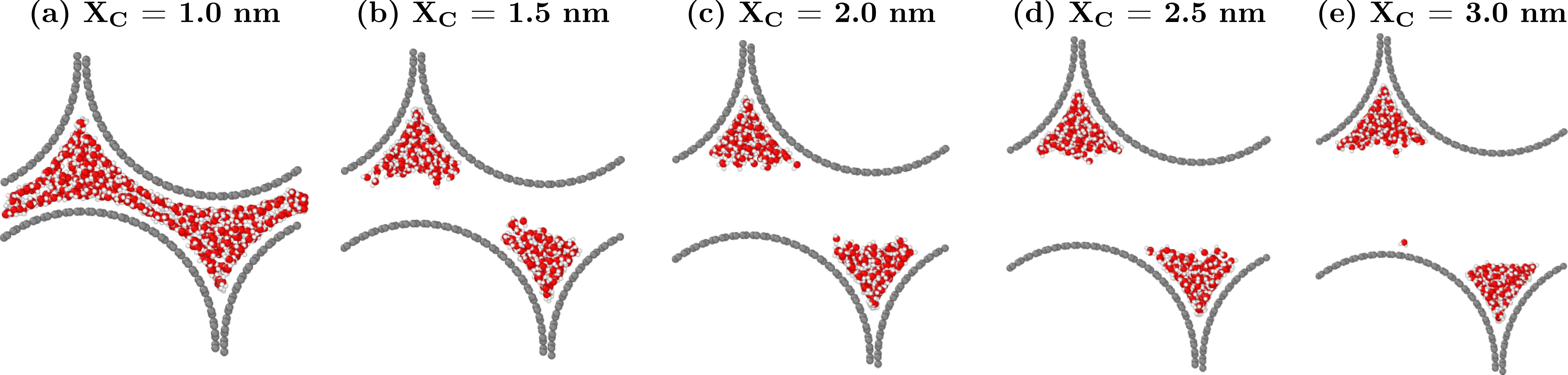}\\
        \includegraphics[width=6.4in]{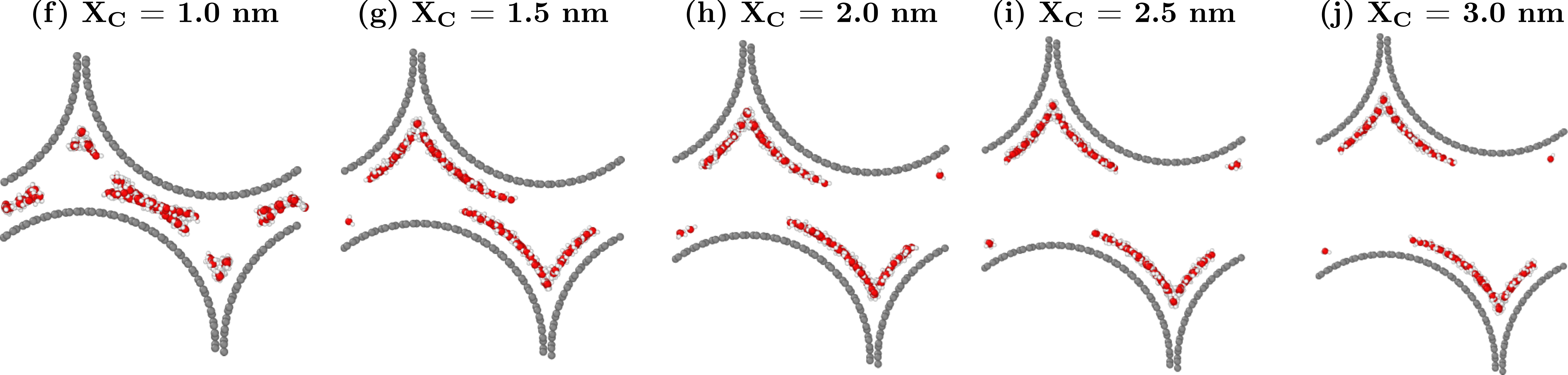}
	\end{center}
	\caption{Two infinite parallel rows of (a)-(e) hydrophobic and (f)-(j) hydrophilic carbon nanotubes (CNT) with a diameter of 5 nm are placed at a distance from each other such that the minimum vertical distance between the rows X$_{c}$ varies from 1.0 nm to 3.0 nm, as indicated. This establishes a 2D nanochannel between the CNT rows where water molecules are inserted at a density of 0.310 molecules/nm$^{3}$.}     
	\label{fig_xc5nm}
\end{figure}

\begin{figure}[H]
	\begin{center}
		\includegraphics[width=6.4in]{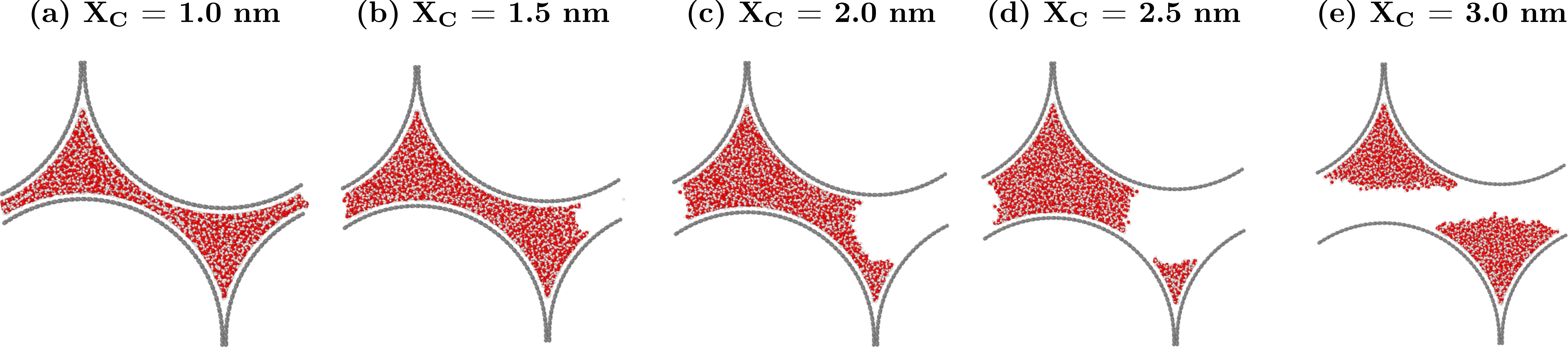}\\
        \includegraphics[width=6.4in]{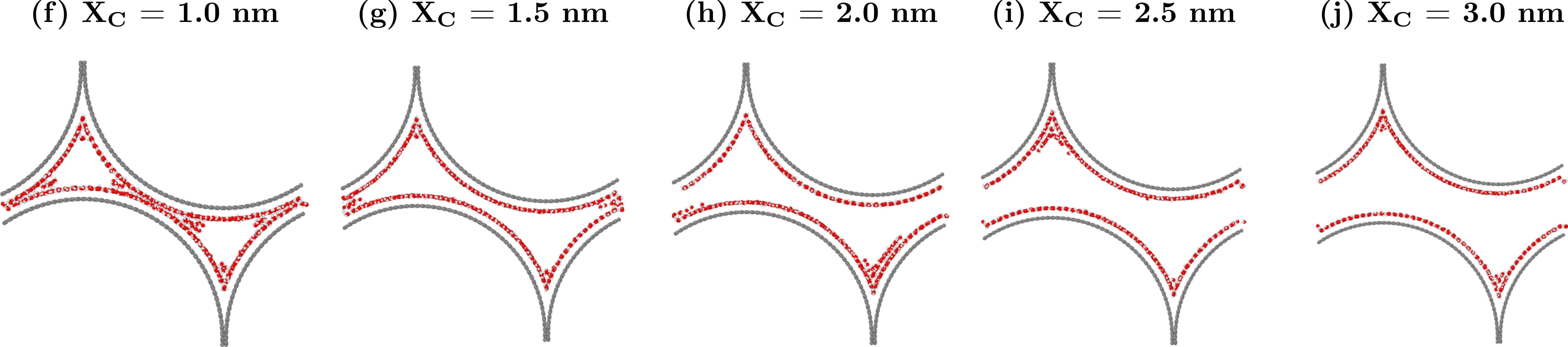}
	\end{center}
	\caption{Same as Fig. \ref{fig_xc5nm}, for nanotubes with diameter of 10 nm. }     
	\label{fig_xc10nm}
\end{figure}

For all values  of nanotube diameters considered, we observed that hydrophilic water-nanotube interaction leads water to form layers at the surface of the nanochannel, as shown in Figures \ref{fig_xc5nm} and \ref{fig_xc10nm} (f)-(j). For hydrophobic water-nanotube interaction, water agglomerates into one dimensional channels at valley regions of the nanochannels. The latter behavior (for hydrophobic interactions) is shown in Figures \ref{fig_xc5nm} and \ref{fig_xc10nm} (a)-(e). For both hydrophilic and hydrophobic systems, water forms connected channels along the nanotubes´ direction (perpendicular to the figure). In the hydrophilic case, water can also maintain a connected channel perpendicular to the nanotubes´ direction as shown in Figure \ref{fig_xc10nm} (f)-(j). 

The above results indicate that, in the hydrophobic case, the dimensionality of the water nanochannel decreases (from two to one) with increasing void region channel width, for a given void water content. This suggests that the critical void region water content, at which a change in the dimensionality of the water channel network occurs, should increase with the width of the void region channel. For the hydrophilic case, hydrophilic surfaces tend to order adjacent water molecules (the interfacial water layer) strongly than hydrophobic surfaces. This ordering is due to the formation of strong, multiple hydrogen bonds between water and the hydrophilic functional groups on the carbon surface. This ordering of water molecules may be linked to strong surface interactions and may, in some cases, select the formation of ice with specific structures near the surface, causing the formed water layer to be linear and continue to follow the geometry of the confining surface due to the strong attractive interaction.

\begin{figure}[H]
	\begin{center}
		\includegraphics[width=6.4in]{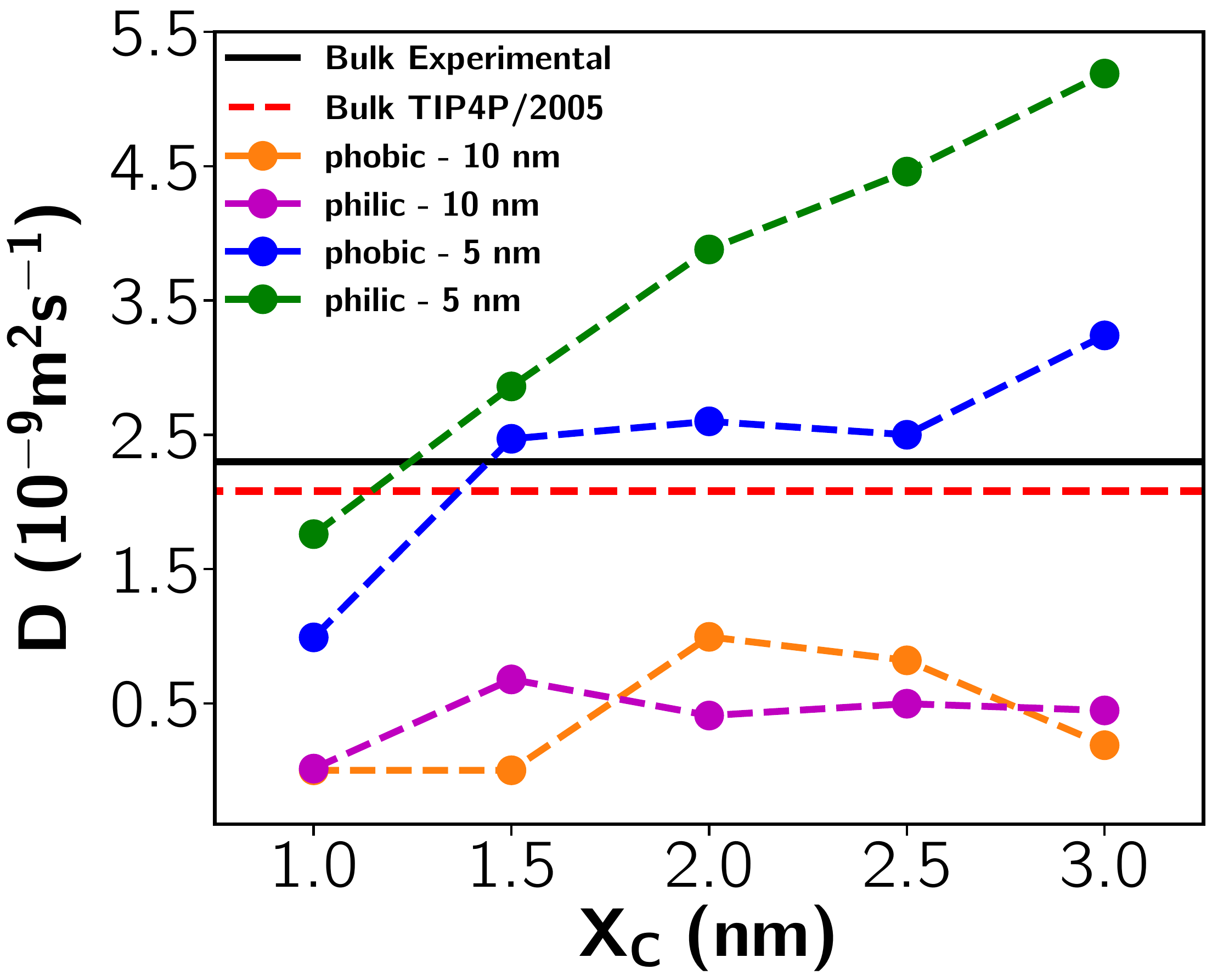}
	\end{center}
	\caption{Axial diffusion coefficient as a function of the minimum vertical distance between rows X$_{c}$ ranging from 1.0 nm to 3.0 nm. The reported values cover nanosheets of 5 nm and 10 nm in diameter for both hydrophobic and hydrophilic interactions. We also present the values for the experimental \cite{harris1980pressure} diffusion coefficient and for the theoretical model TIP4P/2005 \cite{abascal2005general} at 298.15 K.} 
    \label{fig_diff}
\end{figure}

\begin{figure}[H]
	\begin{center}
		\includegraphics[width=6.4in]{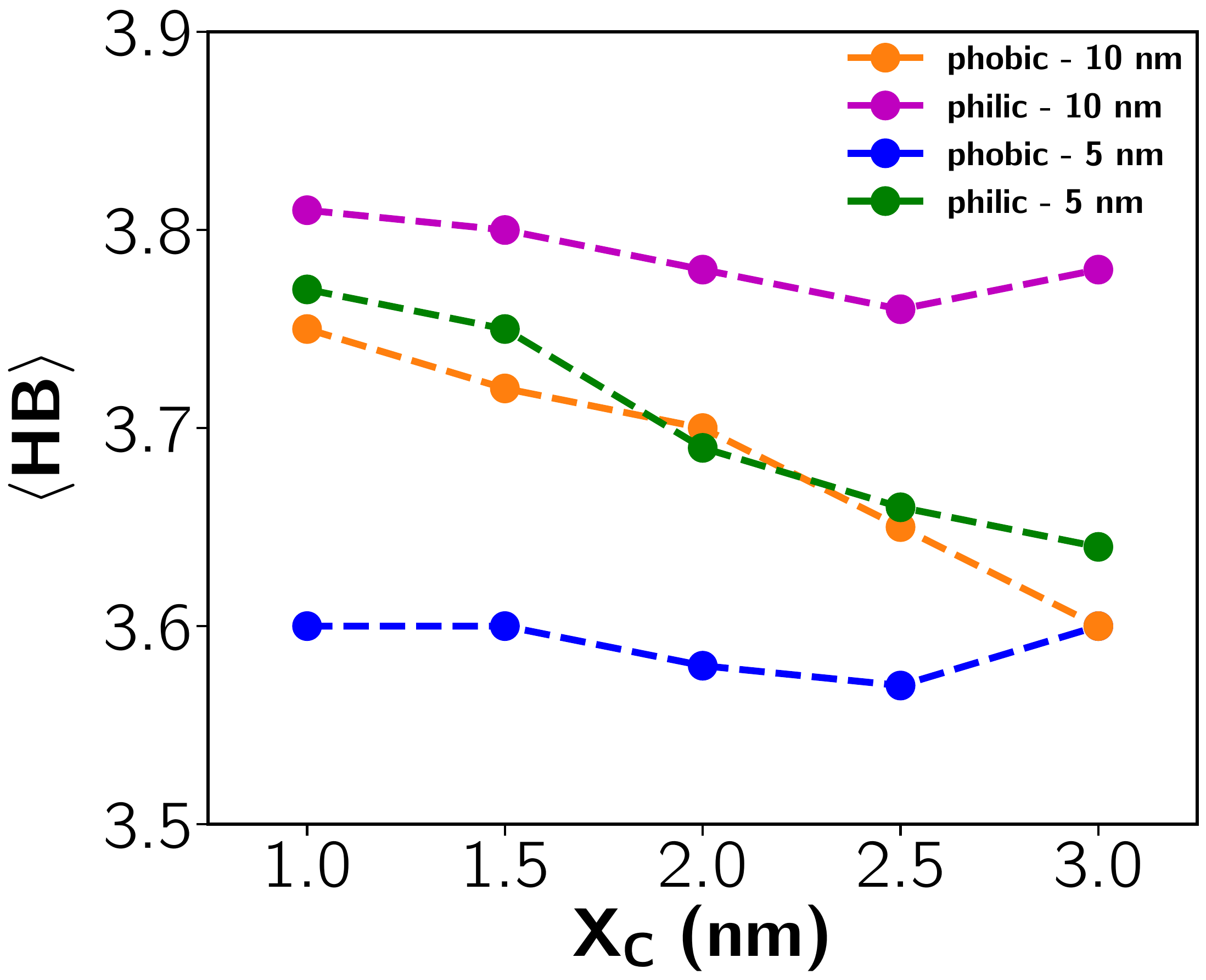}
	\end{center}
	\caption{Average number of hydrogen bonds per water molecule as a function of the minimum vertical distance between rows X$_{c}$ ranging from 1.0 nm to 3.0 nm. The reported values cover nanosheets of 5 nm and 10 nm in diameter for both hydrophobic and hydrophilic interactions.} 
    \label{fig_hbs}
\end{figure}

We now analyze the axial diffusion coefficient (Fig.\ref{fig_diff}) and the hydrogen bonds (Fig.\ref{fig_hbs}) of the systems. Diffusion coefficients for systems with nanotube diameter of 5 nm are always larger than those with nanotube diameter of 10 nm. This phenomenon did not relate to the hydrophobicity or hydrophilicity of the systems. A possible interpretation is that the phenomenon governing the diffusivity of the systems was the confinement length, forcing water molecules to form or break hydrogen bonds, making the interaction of water with the surface secondary to diffusion and, consequently, to hydrogen bonding in these cases of extreme confinement. The diffusion of water in smaller channels is easier than in larger channels because of a phenomenon known as enhanced diffusion, or superdiffusion, by confinement. In systems with dimensions analyzed, water tends to organize itself into single-file structures, thus being forced to organize itself into well-ordered structures, and the movement of water becomes highly cooperative and collective. 

This is clearly depicted in Fig. \ref{fig_diff}, which shows that, for systems with nanotube diameter of 5 nm, the water self-diffusion coefficient surpasses, by a factor of up to twofold, the bulk water value. In systems with larger channels, water remains confined, but the confinement length is large enough to allow the formation of multiple layers of water and a more complex and disordered network of hydrogen bonds, behaving more closely to what we find in bulk water.

\begin{figure}[H]
	\begin{center}
		\includegraphics[width=6.4in]{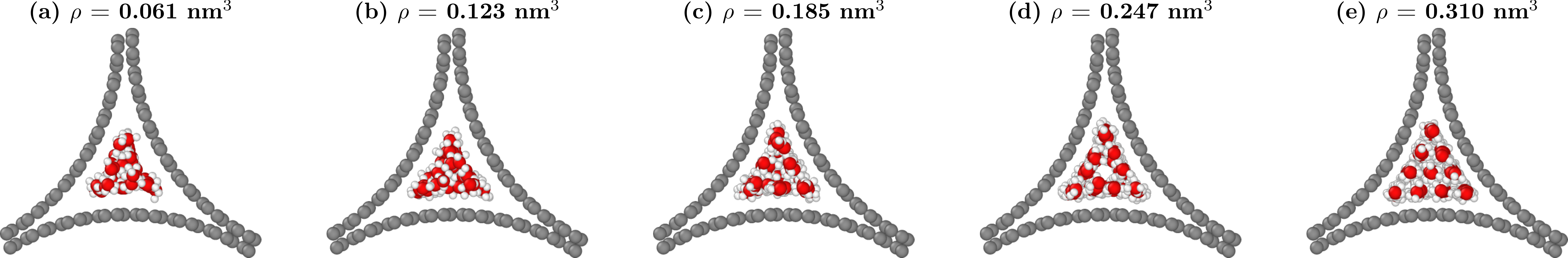}\\
        \includegraphics[width=6.4in]{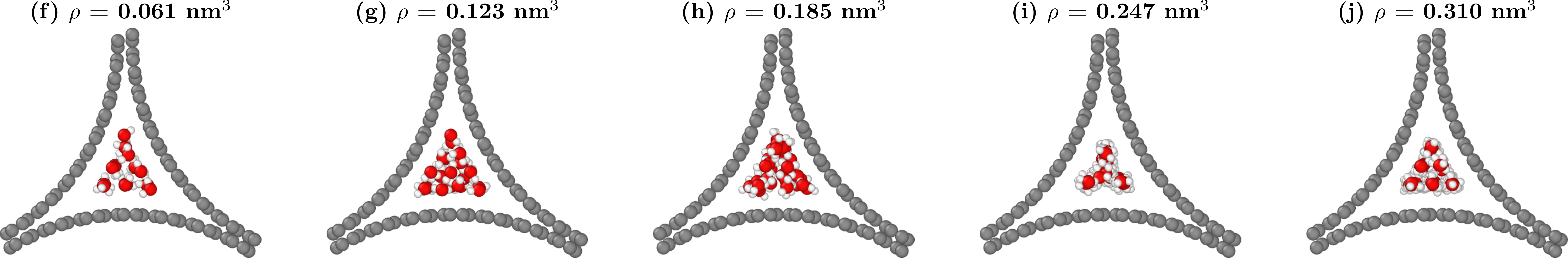}
	\end{center}
	\caption{Snapshots of water confined to interstitial voids at the bulk of nanotube bundles, for both hydrophobic (a)-(e) and hydrophilic (f)-(j) interactions. The diameter of the nanotubes is 5 nm. Water molecules are inserted from a density of 0.061 molecules/nm$^{3}$ (left) up to 0.310 molecules/nm$^{3}$ (right).
    }   	
    \label{fig_dens5nm}
\end{figure}

\begin{figure}[H]
	\begin{center}
		\includegraphics[width=6.4in]{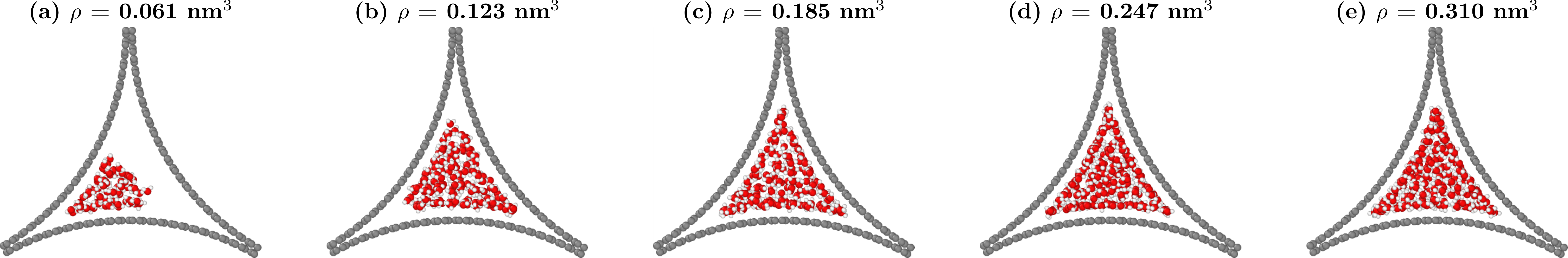}\\
        \includegraphics[width=6.4in]{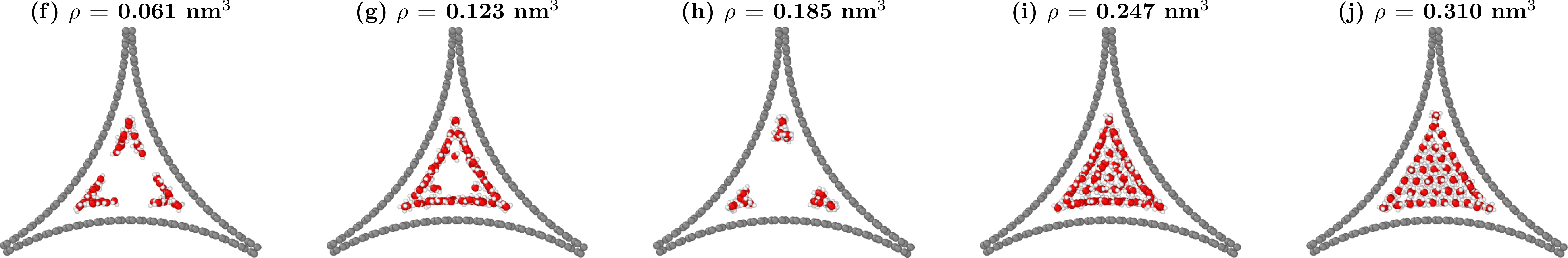}
	\end{center}
	\caption{Same as Fig. \ref{fig_dens5nm}, for a nanotube diameter of 10 nm.}     
	\label{fig_dens10nm}
\end{figure}

\begin{figure}[H]
	\begin{center}
		\includegraphics[width=6.4in]{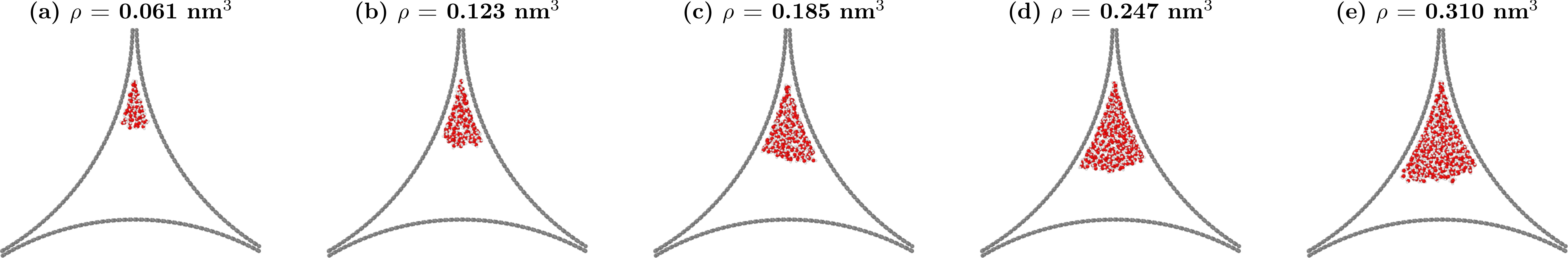}\\
        \includegraphics[width=6.4in]{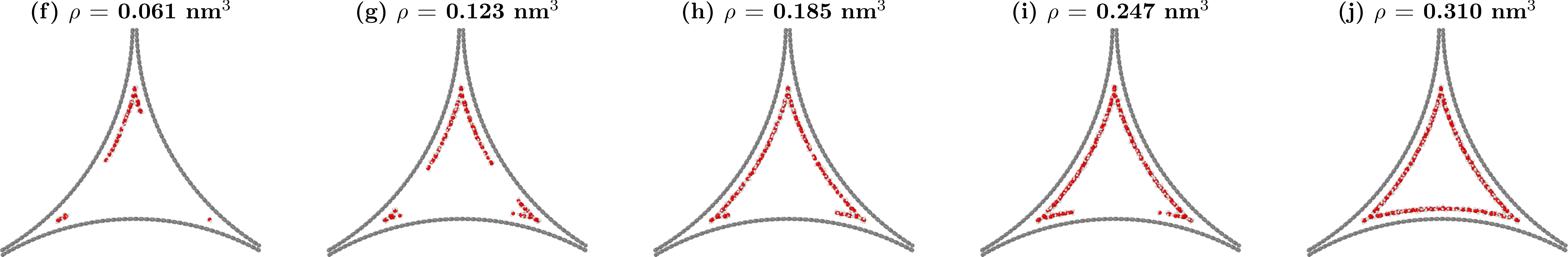}
	\end{center}
	\caption{Same as Fig. \ref{fig_dens5nm}, for a nanotube diameter of 20 nm.}     
	\label{fig_dens20}
\end{figure}

\begin{figure}[H]
	\begin{center}
		\includegraphics[width=6.4in]{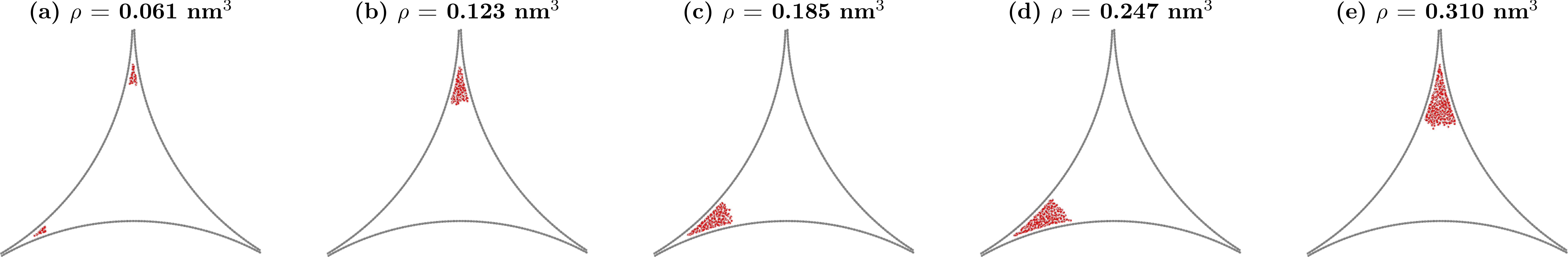}\\
        \includegraphics[width=6.4in]{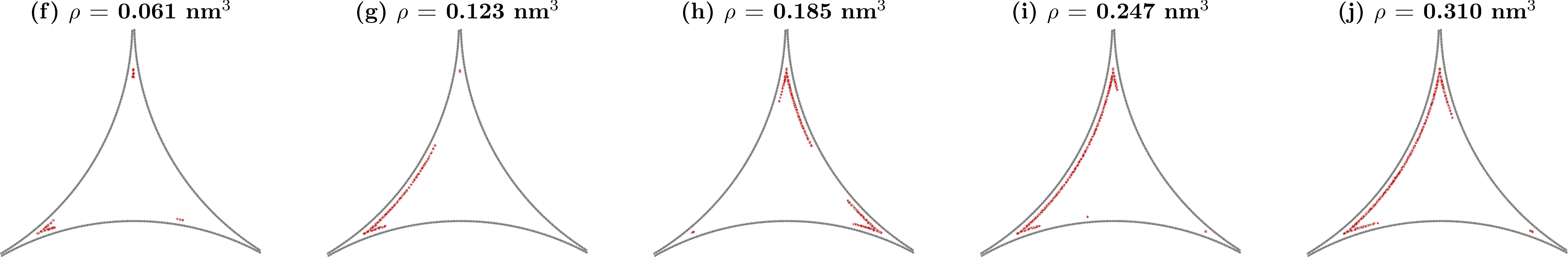}
	\end{center}
	\caption{Same as Fig. \ref{fig_dens5nm}, for a nanotube diameter of 40 nm.}     
	\label{fig_dens40}
\end{figure}

We have so far analyzed properties of water channels at the surface of nanotube bundles. Let us also consider properties of possible water channels at the interior (bulk) region of the bundles, specifically at the interstitial voids between the outer part of the nanotubes. Snapshots of water confined to such interstitial voids are shown in Fig. \ref{fig_dens5nm} for 5 nm diameter nanotubes, in Fig. \ref{fig_dens10nm} for 10 nm diameter nanotubes, in Fig. \ref{fig_dens20} for 20 nm diameter nanotubes, and in Fig. \ref{fig_dens40} for 40 nm diameter nanotubes. The upper panels (a-e) of these figures correspond to hydrophobic water-tube interactions, and the lower panels (f-j) to hydrophilic interactions. In each row of panels, the amount of water increases from left to right. Similarly to the water channels at the surface of the nanotube bundles, hydrophobic interactions leads to the clustering of the water channels at one of the three neighboring ``grooves'' of the voids. A possible driving mechanism for such symmetry breaking is the reduction of water surface area. For hydrophilic interactions, also similarly to the surface channels, water tends to form a nearly uniform wetting layer.

\section{Conclusions}

In this work, we investigate using classical molecular dynamics how water behaves when confined to void regions of nanotube bundles, either between surfaces or at bulk interstitial sites. We focus on the effects of surface chemistry and channel geometry on nanoconfined water structural organization and connectivity. In hydrophobic systems, water molecules tend to gather into isolated one-dimensional channels within the voids, and only at higher water densities do these droplets begin to merge, forming continuous pathways through the structure. In contrast, hydrophilic surfaces favor the formation of ordered and stable water layers that remain connected even under conditions of low humidity because the interaction between water and the surface promotes strong hydrogen bonding at the interface. The degree of confinement also plays a crucial role: when the channels are moderately narrow, the molecules organize into more continuous layers, while very wide or very tight confinements disrupt this connectivity. This leads to large values of waterdiffusionn constant, surpassing the bulk water value for thin channels. In this context, these observations suggest that the collective behavior of percolating water and the transition from an insulating to a conducting regime may emerge more readily in hydrophilic and moderately confined systems, offering a microscopic view of the humidity-dependent conduction in nanoporous systems.

\begin{acknowledgement}
This work is funded by the Brazilian scientific agencies Fundação de Amparo à Pesquisa do Estado da Bahia (FAPESB), Fundação de Amparo à Pesquisa do Estado de Minas Gerais (FAPEMIG), Comissão de Aperfeiçoamento do Pessoal de Ensino Superior (CAPES), Conselho Nacional de Desenvolvimento Científico e Tecnológico (CNPq) and the Brazilian Institute of Science and Technology (INCT) in Carbon Nanomaterials with collaboration and computational support from Universidade Federal da Bahia (UFBA) and the Universidade Federal de Minas Gerais (UFMG). In addition, the authors acknowledge the National Laboratory for Scientific Computing (LNCC/MCTI, Brazil) for providing HPC resources of the SDumont supercomputer, which have contributed to the research results reported within this paper. URL: http://sdumont.lncc.br. Finally, EEM appreciates Edital PRPPG 010/2024 Programa de Apoio a Jovens Professores(as)/Pesquisadores(as) Doutores(as) - JOVEMPESQ Project 24460.
\end{acknowledgement}

\bibliography{achemso-demo}

\end{document}